\documentclass[a4paper,11pt]{article} 
\pdfoutput=1 
\usepackage{jcappub} 
\usepackage[T1]{fontenc} 

\newcommand{\be}{\begin{equation}}
\newcommand{\ee}{\end{equation}}
\newcommand{\simless}{\lower.5ex\hbox{$\; \buildrel < \over \sim\;$}}
\newcommand{\simgreat}{\lower.5ex\hbox{$\; \buildrel > \over \sim\;$}} 
\newcommand{\earth}{ \oplus } 
\newcommand{\lbar}{\langle L \rangle} 
\newcommand{\cross}{ \sigma }

\title{\boldmath Planets in Other Universes: 
Habitability constraints on density 
fluctuations and galactic structure} 

\author[a,b]{Fred C. Adams,}
\author[a]{Katherine R. Coppess,} 
\author[c]{Anthony M. Bloch} 

\affiliation[a]{Physics Department, University of Michigan, Ann Arbor, MI 48109} 
\affiliation[b]{Astronomy Department, University of Michigan, Ann Arbor, MI 48109} 
\affiliation[c]{Mathematics Department, University of Michigan, Ann Arbor, MI 48109} 

\emailAdd{fca@umich.edu} 
\emailAdd{kcoppess@umich.edu} 
\emailAdd{abloch@umich.edu} 

\abstract{Motivated by the possibility that different versions of the  
laws of physics could be realized within other universes, this paper
delineates the galactic structure parameters that allow for habitable
planets and revisits constraints on the amplitude $Q$ of the
primordial density fluctuations. Previous work indicates that large
values of $Q$ lead to galaxies so dense that planetary orbits cannot
survive long enough for life to develop. Small values of $Q$ lead to
delayed star formation, loosely bound galaxies, and compromised heavy
element retention. This work generalizes previous treatments in the
following directions: [A] We consider models for the internal
structure of the galaxies, including a range of stellar densities, and
find the fraction of the resulting galactic real estate that allows
for stable, long-lived planetary orbits. [B] For high velocity
encounters, we perform a large ensemble of numerical simulations to
estimate cross sections for the disruption of planetary orbits due to
interactions with passing stars. [C] We consider the background
radiation fields produced by the galaxies: If a galaxy is too compact,
the night sky seen from a potentially habitable planet can provide
more power than the host star.  [D] One consequence of intense
galactic background radiation fields is that some portion of the
galaxy, denoted as the Galactic Habitable Zone, will provide the right
flux levels to support habitable planets for essentially any planetary
orbit including freely floating bodies (but excluding close-in
planets).  As the value of $Q$ increases, the fraction of stars in a
galaxy that allow for (traditional) habitable planets decreases due to
both orbital disruption and the intense background radiation. However,
the outer parts of the galaxy always allow for habitable planets, so
that the value of $Q$ does not have a well-defined upper limit (due to
scattering or radiation constraints). Moreover, some Galactic
Habitable Zones are large enough to support more potentially habitable
planets than the galaxies found in our universe. These results suggest
that the possibilities for habitability in other universes are somewhat 
more favorable and far more diverse than previously imagined. }

\begin{document}
\maketitle
\flushbottom

\section{Introduction} 
\label{sec:intro} 

One of the most important cosmological parameters is the amplitude $Q$
of the primordial density fluctuations. In our universe, this quantity
is small, $Q \sim 10^{-5}$, as measured by the corresponding
temperature fluctuations in the cosmic microwave background
\citep{cobe,wmap,planck}. The parameter $Q$ sets the initial
conditions for the growth of structure in standard cosmological
models, where dark matter provides most of the matter available for
structure formation and dark energy provides most of the energy
density of the universe \citep{wmap}. 

A key question is why the amplitude $Q$ has this particular small
value. The temperature fluctuations observed in the cosmic background
radiation are thought to arise from quantum fluctuations produced
during the inflationary epoch \citep{guth} in the early universe.
However, the observed smallness of the parameter $Q$ requires a
corresponding small parameter in the scalar field potential of
inflationary models \citep{bardeen,afg}. Other choices of the scalar
field potential generally tend to produce larger fluctuations and
hence larger values of $Q$ (see the discussion of \citep{garriga}).
Given the possible existence of other universes, perhaps as part of a
multiverse, it is reasonable to expect that these other regions of
space-time sample a distribution of values for $Q$.  Previous authors
\citep{tegrees,tegmark} have considered this possibility and placed
bounds on the values of $Q$ that allow for structure formation with
favorable properties.  In general, if the fluctuation amplitude $Q$ is
too small, then cosmological structure has difficulty forming, and the
resulting objects could be too rarified to cool and form stars
\citep{reesost,whiterees,tegrees}; in addition, they could be too
loosely bound to retain the heavy elements produced by stars. On the
other hand, if $Q$ is too large, then galaxies are much denser than
those of our universe; they can be too dense to allow for planetary
orbits to remain unperturbed over the long time scales necessary for
biological evolution to run its course.

The consideration of alternate universes --- in this case with
different values of $Q$ and hence different types of galactic
structures --- is counterfactual in nature. The laws of physics are
described by a collection of fundamental constants, but we currently
have no definitive explanation for why these constants have their
measured values. Similarly, the standard picture of cosmology requires
a corresponding set of cosmological parameters, with specific values
required to match observations, but we have no {\it a priori} theory
to predict these values. One possible -- but only partial --
explanation is that our universe is part of a larger space-time
structure, often called the multiverse \citep{rees,reessix,vilenkin},
where each separate universe represents a different realization of the
laws of physics within the ensemble of possible universes. More
specifically, each separate region of space-time can have its own
values for both the fundamental constants of physics and for the
cosmological parameters. Our universe -- our region of space-time --
thus has one particular realization of these quantities, drawn from
some underlying distribution of parameters.

This possibility motivates the present paper: If the multiverse
concept can provide even a partial explanation for why our universe
has its observed properties, then we must first understand what range
of parameter space allows for a universe in which life (more or less
as we know it) can develop \citep{carr,hogan,aguirre,tegold,tegmark}.
Note that in order to determine the probability of life arising in a
universe, we must not only determine the parameters that allow for
life but also the probability of realizing those parameter values
\citep{barnes}; thus far, we have little understanding of the
underlying probability distributions. As a result, this paper has the
modest goal of constraining what range of galactic properties allows
for the survival of habitable planets, and using the result to
contrain the allowed values of the fluctuation amplitude $Q$.

The question of structure in other universes, including constraints on
the primordial fluctuation amplitudes \citep{tegmark}, has been
considered previously \citep{bartip,tegrees,hogan}. In general, these
prior treatments adopt a global approach, where a large number of
constraints are presented and calculations are carried out in an
order-of-magnitude fashion (see also \citep{weisskopf}). Building on
these previous studies, this paper focuses on the specific issue of
how galactic environments can disrupt potentially habitable solar
systems. This work explores the problem in greater detail by
considering the internal structure of galaxies (including a range of
densities), numerical simulations of the cross sections for orbital
disruption, and the radiation fields provided by the background
galaxy. This paper thus complements previous studies. 

Within our Galaxy, taking inventory of habitable planets has now
become observationally possible. The first planet with an Earth-like
mass that resides in the habitable zone of a main-sequence star has
recently been detected \citep{quintana}. Moreover, projections suggest
that the fraction of all Sun-like stars that have Earth-like planets
in habitable orbits is large, of order 10 percent \citep{etaearth}.
This fraction of potentially habitable planets could be even higher
for smaller stars, which are far more numerous.  Although long
anticipated, a large population of Earth-like planets is rapidly
becoming an observational reality.

In universes with different choices of the primordial fluctuation
amplitude $Q$, the density of stars in the resulting galaxies
varies. For large $Q$, galaxies collapse early, while the background
universe is dense; the resulting galaxies are more compact, so that
the habitability of Earth-like planets can be compromised in two
conceptually different ways: [A] Potentially habitable planets can be
scattered out of their solar systems when the stellar density is
sufficiently high. [B] The planetary surfaces can become too hot due
to the background stellar radiation fields. As an order of magnitude
estimate, both of these mechanisms start to compromise the survival of
habitable planets when galaxies are denser than those of our universe 
by a factor of $\sim3\times10^5$. However, galaxies have internal
structure, so that the stellar density varies enormously within such
an extended stellar system. As a result, we consider galactic models
and find the fraction of the stars within the galaxies that allow for
habitable solar systems. This generalization allows some fraction of
solar systems to survive, even in galaxies with extreme properties.

The rest of this paper is organized as follows. The current paradigm
for cosmological structure formation is briefly outlined in Section
\ref{sec:cosmology}, where we discuss how the properties of dark
matter halos scale with the amplitude $Q$ and then specify models for
galactic structure.  Section \ref{sec:scatter} presents results for
the disruption of solar systems by scattering encounters, including
calculation of interaction cross sections, disruption rates, and the
fraction of solar systems that survive. Section \ref{sec:radiation}
presents the analogous results for disruption due to radiation by
calculating the background radiation fields of the galaxies and
determining the fraction of solar systems that avoid overheating. The
background radiation fields of dense galaxies provide regions where
freely floating planets have the proper temperature to support liquid
water on their surfaces; the extent of these Galactic Habitable Zones
is explored in Section \ref{sec:ghz}. Finally, we conclude in Section
\ref{sec:conclude} with a summary of our results and a discussion of
their implications.

\section{Galaxy Formation and Galactic Structure} 
\label{sec:cosmology} 

\subsection{Formation of Galaxies} 
\label{sec:galform} 

In this section, we use the basic cosmological framework for structure
formation, as developed for our universe, to determine the types of
structure that could form in other universes with alternate values of
the fluctuation amplitude $Q$. Since we want to consider cases that
are markedly different from our own, we use only the most basic
elements of the current theory. To start, we assume that the universe
is spatially flat and that the matter content is dominated by
collisionless dark matter. In this context, the basic elements of
structure formation with varying $Q$ have been outlined previously for
universes containing only dark matter \citep{tegrees} and including
dark energy \citep{tegmark}. The derivation presented below follows
these earlier papers (see \citep{tegrees,tegmark} for further detail).

Density fluctuations begin to grow at the end of the radiation
dominated era and become nonlinear after the overdensity reaches a
critical value. For the case of a top-hat density perturbation with
spherical symmetry, linear perturbation theory \citep{gungott}
predicts that nonlinear collapse ensues at an overdensity of $\sim1.7$
(see also \citep{schecter}). At the epoch of matter/radiation
equality, the temperature $T_{\rm eq}$ of the universe is given by 
\be 
T_{\rm eq} = \eta m_P {\Omega_M \over \Omega_{\rm b}} \,,
\ee
where $m_P$ is the proton mass, $\eta$ is the baryon-to-photon ratio,
$\Omega_M$ is the dark matter energy density relative to the critical
density, and $\Omega_{\rm b}$ is the corresponding energy density in
baryons \citep{kolbturn,tegmark}.  Note that we are working in units
where $c=\hbar=k=1$, so that $G=M_{\rm pl}^{-2}$ (i.e., $M_{\rm pl}$
is the Planck mass). Notice also that $T_{\rm eq}$ is essentially the
mass in non-relativistic matter per photon.\footnote[2]{Finally, note 
that we ignore the contribution of neutrinos in defining $T_{\rm eq}$.} 
The age of the universe at equality is then given by  
\be
t_{\rm eq} = {1 \over 8} \left( {3 \over \pi} \right)^{1/2} 
{M_{\rm pl} \over a_R^{1/2} T_{\rm eq}^2} \approx 
{M_{\rm pl} \Omega_{\rm b}^2 \over 8 a_R^{1/2} 
(\eta m_P \Omega_M)^2} \,.
\ee 
The mass scale of the horizon at the epoch of matter domination plays
an important role and is given by 
\be
M_{\rm eq} \approx {1 \over 64} {M_{\rm pl}^3 \over T_{\rm eq}^2} 
\approx 4 \times 10^{71} {\rm GeV} \approx 4 \times 10^{14} M_\odot \,, 
\ee
where we have used $\eta = 10^{-9}$ and $\Omega_M/\Omega_{\rm b}=6$, values
appropriate for our universe, to obtain the numerical estimates. 
Fluctuations cannot grow before the epoch of matter domination, 
so that all mass scales $M < M_{\rm eq}$ start with roughly equivalent 
fluctuation amplitudes and grow to become virialized at the later 
time given by 
\be
t_{\rm vir} \approx t_{\rm eq} Q^{-3/2} f_{\rm vir} \,, 
\label{tvirial} 
\ee
where the dimensionless parameter $f_{\rm vir}\simless1$ is a slowly
varying function of the halo mass \citep{tegrees}.  In actuality, the
virialization time (\ref{tvirial}) becomes independent of $M$ only in
the limit of small mass scales; by using this approximation, we are
thus encapsulating the mass dependence of the virialization time into
the parameter $f_{\rm vir}$. Galaxies with smaller masses collapse to
somewhat higher densities because they collapse earlier, when the
universe was denser, and the density of the background universe at the
time of collapse is a controlling factor in the problem (see Section
2.2 and especially Figure 1 of reference \citep{tegrees} for further
discussion). 

After a galactic halo structure collapses, it has a characteristic
density given by a factor of $f_{\rm c} \approx 18 \pi^2$ times the density
of the background universe at time $t_{\rm vir}$, i.e., 
\be
\rho_{\rm c} = f_{\rm c} \rho(t_{\rm vir}) = 18 \pi^2 \rho_{\rm eq} 
\left({t_{\rm eq} \over t_{\rm vir}}\right)^2 = 
18 \pi^2 (2 a_R T_{\rm eq}^4) Q^3 f_{\rm vir}^{-2} \,. 
\label{rhoc}  
\ee
Inserting typical values, we obtain 
\be
\rho_{\rm c} \approx 5.5 \times 10^{-14} \,{\rm g}\,\,{\rm cm}^{-3} 
\,Q^3 f_{\rm vir}^{-2} = 8.2 \times 10^8 M_\odot 
\,{\rm pc}^{-3} \, Q^3 f_{\rm vir}^{-2} \,. 
\label{rhoctwo}
\ee
For mass scales $M < M_{\rm eq}$, the characteristic densities of the
collapsed structures are roughly comparable and given by equations
(\ref{rhoc}) and (\ref{rhoctwo}). The density increases slowly with
decreasing galactic mass due to the factor $f_{\rm vir}$. Since we are
interested in constraints produced by the densest galactic structures,
and since galaxies in our universe have masses $M < M_{\rm eq}$, this
paper focuses on this ``low-mass'' regime. 

To obtain numerical values for equation (\ref{rhoctwo}), we have used
the baryon-to-photon ratio $\eta=10^{-9}$ and the mass density ratio
$\Omega_M/\Omega_{\rm b}=6$, although these quantities are expected
to vary from universe to universe. As a result, one should keep in
mind that the characteristic density scales according to the relation 
$\rho_{\rm c} \propto Q^3T_{\rm eq}^4$. For ease of presentation, the 
results of this paper are given as constraints on value of $Q$, but
the results more generally apply to the composite parameter 
$QT_{\rm eq}^{4/3}$. In other words, larger values of the fluctuation
amplitude $Q$ can be offset by invoking alternate values of
$(\eta,\Omega_M,\Omega_{\rm b})$ that lead to smaller values of
$T_{\rm eq}$ (keeping the composite parameter constant). On the other
hand, inflationary scenarios tend to produce a wide range of possible
fluctuation amplitudes, often larger than those in our universe, so it
makes sense to use $Q$ as the primary variable.

For completeness, we note that for mass scales $M>M_{\rm eq}$ collapse
occurs later, so that these large structures become nonlinear (as 
density perturbations) at later times given roughly by 
\be
t \approx t_{\rm eq} Q^{-3/2} f_{\rm vir} (M/M_{\rm eq}) \,.
\ee 
Alternatively, the mass scales of these larger structures are 
a fraction of the horizon mass, 
\be
M \sim Q^{3/2} M_{\rm hor} \sim {4 \pi \over 3} \rho r_{\rm hor}^3 \,, 
\ee
where $r_{\rm hor}$ is the horizon scale at the time when the
perturbations become nonlinear. The corresponding length scale $r$ is
given by 
\be
r \sim Q^{1/2} r_{\rm hor} \,. 
\ee
These more massive structures have lower characteristic densities by a
factor of $\sim(M_{\rm eq}/M)^2$. As a result, the lower-mass regime
considered above produces galaxies that are more disruptive to their
constituent solar systems.

The considerations outlined above apply to the initial formation of
cosmological structures. As the universe in question evolves,
structures on all scales tend to interact and merge, thereby creating
a complicated network of interactions. Merging continues until the
dark energy component of the universe (if nonzero) dominates and
effectively freezes out further structure formation
\citep{nagamine,busha2005,busha2007,tegmark}. While operative, the
merging process tends to produce ever-larger structures and acts to
redistribute the mass --- both the internal structure of individual
halos and the mass distribution of the ensemble of halos. In general,
merging structures result in more massive structures with lower
density. Since we are primarily interested in the densest structures,
however, we focus on halos with masses $M{\simless}M_{\rm eq}$.

\subsection{Structure of Galaxies} 
\label{sec:galstructure} 

Numerical simulations of cosmological structure formation indicate
that dark matter halos of galaxies and clusters approach a nearly
universal form. The benchmark study of this convergence \citep{nfw}
showed that the density distribution of the halos assume what is now
called the NFW profile, which can be written as 
\be
\rho = {\rho_0 \over \xi (1 + \xi)^2} 
\qquad {\rm where} \qquad \xi = {r \over r_0} \,, 
\label{rhonfw} 
\ee
where $r_0$ is the scale length of the system. This simple form for
the density distribution cannot continue out to arbitrarily large
radii (as the enclosed mass would diverge). However, a number of
additional effects act to provide halo edges. At the present
cosmological epoch, halos with this density distribution meet up with
their neighboring halos, which provide an effective outer boundary. 
At later times, however, the accelerating cosmic expansion leads to
the isolation of halos, which approach an asymptotic form
\cite{busha2005,busha2007} with a somewhat steeper density
distribution with the form of a Hernquist \citep{hernquist} profile, 
\be 
\rho = {\rho_0   \over \xi (1+\xi)^3} 
\qquad {\rm where} \qquad \xi = {r \over r_0} \,.
\label{hernquist} 
\ee
These same simulations show that a truncation radius develops in above
density distribution, where the asymptotic value $r_T \approx 4.6 r_{200}$
(see Figure 3 of \cite{busha2005}; see also \cite{kravtsov}). 

In this work, we assume that dark matter halos have the form given
by equation (\ref{hernquist}). For simplicity, we also assume that the
stellar component of the galaxies have the same form, albeit with a
more concentrated configuration. In other words, we also use the
Hernquist profile to specify the distibution of stars for a model 
galaxy. We can then calculate the density as a function of radius. At
sufficiently large distances, the density will decrease enough so that
planets can survive the disruptive effects of both scattering and
radiation.

The (crude) theory of structure formation outlined above shows that 
for a universe with a given value of $Q$, all halos with moderately 
low mass $M < M_{\rm eq}$ will virialize at roughly similar times and 
have densities $\rho_{\rm c}$ given by equation (\ref{rhoc}) after collapse. 
Here we make the identification that 
\be
\rho_{\rm dm} = \rho_{\rm c} \,, 
\ee
where $\rho_{\rm dm}=\rho_0$ is the density scale appearing in the density
distribution (\ref{hernquist}) for the dark matter. For a given total 
halo mass $M$, the corresponding length scale $r_{\rm dm}$ for the halo is
given by 
\be
r_{\rm dm} = \left( {M \over 2\pi \rho_{\rm c}} \right)^{1/3} = 
\left( {M \over M_{\rm eq}} \right)^{1/3} {M_{\rm pl} \over 8\pi T_{\rm eq}^2} 
\left( {f_{\rm vir}^2 \over 9 a_R} \right)^{1/3} Q^{-1} \,. 
\label{rdark} 
\ee

With the parameters of the dark matter halo specified, we need to
determine the corresponding parameters for the baryonic component.
Since gas can dissipate energy, one expects the baryons to collapse
further and attain a more centrally concentrated configuration.
Because the ratio of the total mass in baryons to that in dark matter
is determined by $\Omega_M/\Omega_{\rm b}$, only one additional
parameter needs to be specified. We take this parameter to be the
ratio $R_\rho$ of the density scales of the two components, i.e., 
\be
R_\rho \equiv {\rho_{\rm b} \over \rho_{\rm dm}} \,. 
\label{rratio} 
\ee
Keep in mind that $\rho_{\rm b}$ (respectively, $\rho_{\rm dm}$) is
the density scale appearing in the baryonic (dark matter) density
profile, both of which are assumed to have the Hernquist form given 
by equation (\ref{hernquist}). The scale length $r_{\rm b}$ for the
baryonic component is then given by 
\be
r_{\rm b} = r_{\rm dm} \left( {\Omega_{\rm b} \over \Omega_{\rm M}} 
{\rho_{\rm dm} \over \rho_{\rm b}} \right)^{1/3} = r_{\rm dm} 
\left( {\Omega_{\rm b} \over \Omega_{\rm M} R_\rho} \right)^{1/3} \,. 
\label{rbary} 
\ee
If the star formation rate is uniform across the galaxy, the 
number density $n_\ast(\xi)$ of stars also has the form 
\be
n_\ast = {n_0 \over \xi (1+\xi)^3} \qquad {\rm where} \qquad 
\xi = {r \over r_{\rm b}} \,.
\label{numhern} 
\ee
The total number of stars in the (spherical) 
galaxy is given by 
\be
N_\ast = 2\pi r_{\rm b}^3 n_0 \,, 
\ee
where the stellar density scale $n_0$ is given by 
\be
n_0 = 
{\epsilon_{\rm sf}\,\rho_{\rm b} \over \langle M_\ast \rangle}\,,
\label{nzero} 
\ee
where $\epsilon_{\rm sf}$ is the star formation efficiency of the
galaxy (the fraction of the baryonic mass within the galaxy that has
been processed into stars) and $\langle M_\ast \rangle$ is the mean
stellar mass.

This treatment for the baryonic component makes several assumptions:
First we assume that the gas can cool and successfully make stars. In
order to cool promptly, on time scales comparable to the age of the
universe at the epoch of formation, the gas must be sufficiently
dense. In this context, previous work shows that for $Q \simgreat$
$3\times10^{-5}$, the gas can cool for essentially all mass scales of
interest \citep{tegrees}. As a result, the cooling criterion is
expected to be satisfied for most galaxies in the large-$Q$ universes
considered in this paper.

In our universe, during galaxy formation the gas often falls inward to
form disk-like structures, rather than the simple spherical profiles
considered here. In addition to providing a useful working model for
galactic structure, the quasi-spherical density profiles invoked here
are likely to arise for a number of reasons: The characteristic length
scales for galactic structure decrease with increasing $Q$. In our
galaxy, and most other spirals in our universe, the inner regions (on
small length scales) display nearly spherical geometry in the form of
galactic bulges. In fact, the Hernquist profile (equation
[\ref{hernquist}]), which characterizes the asymptotic form of dark
matter halos \cite{busha2005}, was originally developed as a model for
galactic bulges \citep{hernquist}. This form for the density profile
also arises in a wide variety of other cold collapse scenarios (e.g.,
\citep{cannizzo,boily}). If cooling and star formation proceeds
rapidly, galaxies are expected to resemble scaled-up bulges, rather
than disks. Even if they form initially, disk structures can relax
dynamically by scattering stars and adjusting their structure into
more rounded configurations \cite{bintrem}. For our galaxy, the disk
will relax over a long time scale of order $\sim10^{19}$ yr
\citep{dyson,al1997}. For the densest galaxies under consideration,
the relaxation time can become shorter than the typical time for
habitability (usually taken to be $\sim1$ Gyr), but will often be
longer. However, a particular kind of orbit instability --- namely
resonant motion perpendicular to the galactic disk --- can force stars
(or parcels of gas) to leave their orbital planes in even mildly
triaxial systems \citep{binney81}. In many cases, the time scale for
this type of instability to alter the phase space of a stellar system
is much shorter than the time scale for two-body relaxation
\citep{aplus}. Finally, merging of smaller structures plays an
important role in determining the form of bulges and ellipticals in
our universe, and mergers could be even more important in the high-$Q$
universes of interest here. This paper thus assumes that the
combination of rapid collapse, mergers, orbit instability, and
dynamical relaxation leads to rounded galactic structures.

\section{Scattering Constraints} 
\label{sec:scatter} 

In this section we estimate the disruption probability for solar
systems in the idealized galactic models described above.  The rate
$\Gamma$ of scattering encounters between a solar system and passing
field stars is given by 
\be
\Gamma = n_\ast \cross v \,,
\ee
where $n_\ast$ is the stellar density, $\cross$ is the interaction
cross section, and $v$ is the relative speed between the solar system
in question and passing stars.

We start with an order of magnitude estimate: In our Galaxy today
\citep{binmer}, the stellar density $n_\ast \approx 0.2$ pc$^{-3}$ and
the random encounter speed $v \approx 40$ km s$^{-1}$ = 40 pc/Myr. At
these speeds, the cross section for disrupting the orbit of Earth is
approximately $\cross \approx 50$ AU$^2$ = 1.25 $\times 10^{-9}$
pc$^2$ \citep{frozen}, where disruption corresponds to an increase in
eccentricity to values above $e>0.5$ (with this eccentricity, the
stellar flux varies by a factor of 9, nearly an order of magnitude,
over the orbit).  With these values, the benchmark interaction rate
for the disruption of our Solar System becomes 
\be
\Gamma = 10^{-8} {\rm Myr}^{-1} = 10^{-5} {\rm Gyr}^{-1} \,.
\ee 
Although the time scale required for life to develop is not known, 4.6
Gyr is apparently long enough and the value of $t_{\rm c}$ = 1 Gyr is often
used as a standard time scale required for habitability
\citep{lunine,scharf}. Adopting this value, we would have to increase
the stellar density by a factor of $\sim10^5$ in order for the
interaction rate to be high enough to compromise habitability (keeping
other parameters fixed).

\subsection{Interaction Cross Sections} 
\label{sec:cross} 

In this work, we need to calculate the cross sections for disruption
of planetary orbits due to interactions between a solar system and
passing stars. Previous work has explored these cross sections for the
lower speeds appropriate for star forming cluster environments
\citep{al2001,gdawg} and for field stars in our galaxy \citep{frozen}.
For the dense, highly interactive galaxies of interest in this work,
we need the disruption cross sections for higher encounter speeds. 

For the sake of definiteness, we focus on the case of a single planet
system interacting with a single passing star\footnote[2]{For the dense
galaxies of interest, binary systems act like two separate single stars
because the encounter speeds are much larger than binary orbit speeds
\citep{gdawg}.} and take the system to be an analog of the Earth-Sun
system: The host star mass $M_{\rm host}=1M_\odot$, the planet mass
$M_P=1M_\earth$, and the initial planetary orbit is circular ($e=0$)
with semimajor axis $a=1$ AU.  Even in this simplified setting, the
interactions must be described by a large number of additional
variables, including the mass $M_\ast$ of the passing star, the impact
parameter $b$ of the encounter, the relative velocity $v_r$ of the two
stars at infinite separation, two angles $(\theta,\phi)$ that specify
the orientation of the solar system with respect to velocity vector of
the incoming star, and the phase angle of the planetary orbit at the
start of the interaction.  We approach this problem using a Monte
Carlo scheme to sample the distributions of all of the aforementioned
variables. The stellar masses of the passing stars are drawn from the
initial mass function, which is assumed to have the same form as that
found in our universe.  The angles are drawn randomly such that the
passing star is equally likely to approach from any direction and the
phase angle of the orbit is uniform-random. The relative velocity is
drawn from a Maxwell-Bolzmann type distribution of the form $df/dv$
$\propto v^2 \exp[-v^2/v_{\rm b}^2]$. The velocity scale $v_{\rm b}$
of the distribution characterizes velocity dispersion of the
background galaxy, where the mean value $\langle{v}\rangle$ = 
$2v_{\rm b}/\sqrt{\pi}\approx1.128v_{\rm b}$. Note that the galaxy 
could, in principle, have additional rotational velocities. 

With the variables chosen for a given interaction, the Newtonian
equations of motion are integrated numerically using a Bulirsch-Stoer
algorithm. Note that the integrations must be carried out to high
accuracy: We need to resolve the orbital elements of the planet 
at the end of the simulation; the planet carries energy of order  
$M_P v_{\rm orb}^2$, whereas the energy of the interaction is much
larger, of order $M_\ast v_{\rm b}^2$. The orbital speed $v_{\rm orb}
\sim 30$ km s$^{-1}$, but we need to consider relative speeds as large
as $v_{\rm b}\sim10^4$ km s$^{-1}$; the ratio of these energies is
thus $\sim 10^{-10}$, so the numerical scheme must conserve energy at
an even greater level of precision. For these simulations, the time 
step is reduced accordingly so that this criterion is met (see also 
\citep{frozen,al2001,gdawg} for further discussion).

For each given velocity dispersion of the background galaxy, we have
performed an ensemble of numerical simulations of interactions between
single planet solar systems and passing stars. The results are then
used to find the cross sections for a given change in the orbital
elements of the Earth-analog planet \citep{al2001,gdawg,frozen}. This
paper reports the results from scattering interactions using five
choices of the velocity $v_{\rm b}$ that sets the scale of the velocity
distribution; specifically we use $\log_{\rm 10}(v_{\rm b}/1{\rm km\,\,s}^{-1})$ 
= 2, 2.5, 3, 3.5, and 4 (i.e., the velocity scale $v_{\rm b}$ is distributed 
evenly in the logarithm from 100 to $10^4$ km s$^{-1}$). The total
number of numerical experiments was of order one million, with a
greater number of realizations required for larger velocities. With
the simulations completed, we then post-process the results to
determine the probabilities and the corresponding cross sections for
any particular outcome to occur (details of this procedure are given
in \citep{frozen,al2001,gdawg}).

\begin{figure}[tbp]
\centering 
\includegraphics[width=.90\textwidth,trim=0 150 0 150,clip]{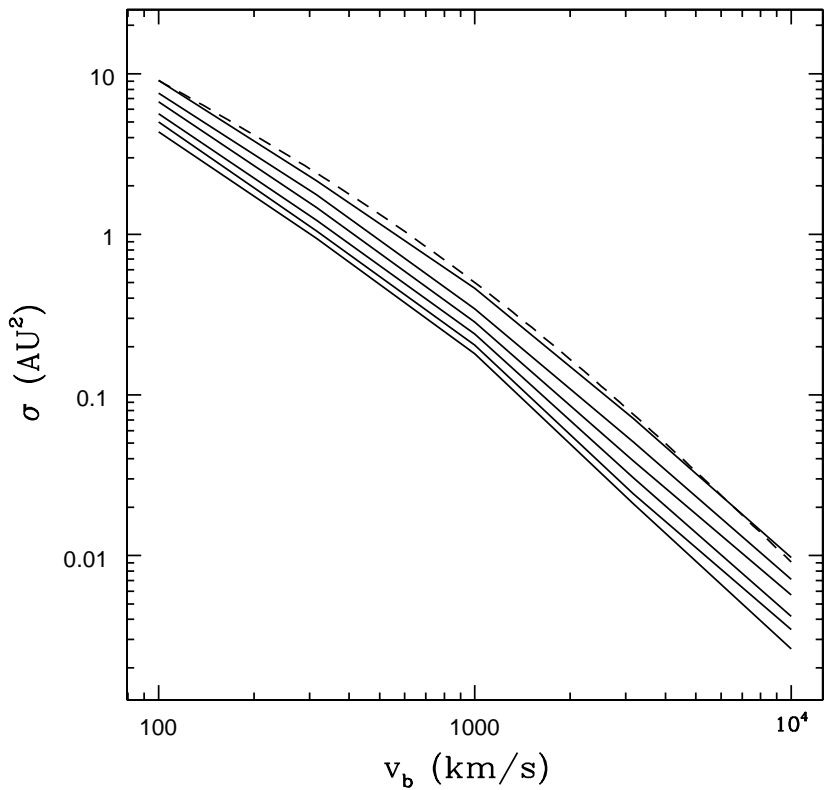}
\caption{\label{fig:csection} Cross sections for the disruption of a 
single-planet solar system due to passing stars. Cross sections are 
shown for increasing the orbital eccentricity to values of 0.5 -- 1.0, 
or greater, from top to bottom, where post-encounter values $e \ge 1$
correspond to planetary ejection. The dashed curves shows a simple 
analytic fit to the cross section for obtaining $e > 0.5$ (see text).} 
\end{figure}

The resulting cross sections are shown in Figure \ref{fig:csection}
for galactic velocity dispersions in the range $v_{\rm b}$ = $10^2-10^4$ km
s$^{-1}$.  Even with the large number of realizations, the Monte Carlo
sampling errors for these cross sections are about 5\%. We also note
that these cross sections are in good agreement with extrapolations
from previous simulations performed at lower impact speeds
\citep{frozen,gdawg}.  The figure shows the cross sections for
increasing the eccentricity of the planet from its starting value
($e=0$) to a post-encounter value of $e$ = 0.5 -- 1, where the $e=1$
final states represent ejections of the planet (including both
stripping the planet from its bound orbit and driving the eccentricity
to unity so that the planet is accreted by its host star).  For
purposes of this paper, we consider $e > 0.5$ to represent sufficient
disruption for compromising planetary habitability and thus adopt the
upper curve in Figure \ref{fig:csection} as the destruction cross
section.

Figure \ref{fig:csection} shows that the disruption cross sections are
steeply decreasing functions of the velocity dispersion (set by $v_{\rm b}$)
of the galaxy. The computed cross section for increasing the
eccentricity to $e>0.5$ (the case of interest) is comparable to the
geometric cross section of the orbit ($A={\pi}a^2\approx3.1$ AU$^2$)
for velocity dispersion $v\sim{v_0}$ = 1000 km s$^{-1}$. As a result,
to a good approximation, the cross section for disruption
$\cross_{\rm dis}$ can be fit with a simple function of the form 
\be
\cross_{\rm dis} \approx { \cross_0 \over u (1 + u) } 
\qquad {\rm where} \qquad u \equiv {v \over v_0} = 
{v \over 1000\,\,{\rm km\,\,s}^{-1} } \,,
\label{crossfit} 
\ee
where $\cross_0$ = 1 AU$^2$.  This expression captures the numerically
discovered behavior that the cross section falls more steeply when the
velocity dispersion $v\simgreat v_0 = 1000$ km s$^{-1}$ and approaches the
form $\cross\propto{v}^{-2}$ in the limit $v\to\infty$. This latter
asymptotic form can be derived analytically using the impulse
approximation and assuming that the passing star acts on the planet
alone (so that the interaction with the host star is negligible). This
latter assumption only holds in the limit where the cross section is
much smaller than the geometric cross section and hence in the high
speed limit.

\subsection{Disruption Rates}  
\label{sec:disrupt} 

With the interaction cross sections and the structure of the galaxy
specified, we consider the disruption rates for solar systems residing
at varying locations. Within the context of this galactic model, some
portion of the the inner galaxy will always be too dense for solar
systems to survive (because $n_\ast \propto 1/\xi$). Similarly, some
solar systems can survive in any galaxy provided that the system
resides at a sufficiently distant orbit. The goal is thus to determine
that fraction of the stellar population that could in princple harbor
habitable planets. 

The disruption rate for solar systems is given as a function of
position within the galaxy by the expression 
\be
\Gamma = {n_0 \cross v \over \xi (1+\xi)^3} \,,
\ee
and the corresponding requirement that solar systems 
survive for a given time $t_{\rm c}$ takes the form 
\be
n_0 \cross v t_{\rm c} < \xi (1+\xi)^3  \,,
\ee
where $\xi=r/r_{\rm b}$. As outlined above, this paper adopts the
characteristic time scale $t_{\rm c}$ = 1 Gyr \citep{lunine,scharf}, which
is a resonable estimate for the time required for biological evolution
to occur.  The reader can scale the results for different choices of
this time scale. Using the fitting function for the interaction cross
section from equation [\ref{crossfit}], this constraint takes the form 
\be
n_0 \cross_0 v_0 t_{\rm c} < \xi (1+\xi)^3 (1 + u) \,,
\label{scatcon1} 
\ee 
where the scaled speed $u$ is a function of galactic position. To
evaluate the speed $u$, we use circular orbits (of the solar system,
around the galaxy) in the potential of a dark matter halo with the 
density distribution of equation (\ref{hernquist}), and find 
\be
u = \left( {G M \over r_{\rm dm} v_0^2} \right)^{1/2} 
{ (\xi r_{\rm b}/r_{\rm dm})^{1/2} \over 1 + \xi r_{\rm b}/r_{\rm dm}} \,.
\label{scatcon2} 
\ee 
The factors of $r_{\rm b}/r_{\rm dm}$ arise because the orbital speed is
determined by all of the mass in the galaxy, which is dominated by
dark matter, whereas the number density of stars (and hence $\xi$)
depends on the baryonic component, which can have a different scale
length. Here $M$ is the total mass of the galaxy.  The interaction
rate is given by the relative speeds of the solar system with respect
to passing stars. The stars are expected to have random velocities
superimposed on their overall orbital motion, where these random
speeds are a fraction of the circular speed. Since both stars are
moving, however, the relative speed is larger than the random speed.
For purposes of this paper, we assume these effects cancel and use
equation (\ref{scatcon2}) to determine the relative speeds. In the
future, a more detailed treatment of the relative velocities should be
developed, along with considerations of the orbits through the galaxy
\citep{ab2005}.

\subsection{Survival Fractions for Scattering} 
\label{sec:survive}  

Equations (\ref{scatcon1}) and (\ref{scatcon2}) define the constraint
that must be met in order for a solar system to survive over the time
$t_{\rm c}$ (taken here to be 1 Gyr).  To evaluate this constraint for
a given galaxy mass $M$, we need to determine the stellar density
scale $n_0$, and length scales $r_{\rm dm}$ and $r_{\rm b}$ for the
dark matter halo and the baryonic component, respectively. Equation
(\ref{nzero}) specifies $n_0$, where we take the star formation
efficiency $\epsilon_{\rm sf}$ = 0.5, the mean stellar mass 
$\langle M_\ast \rangle = 0.5 M_\odot$, and the ratio $R_\rho=10$.
This latter choice reflects the fact that baryons dissipate energy and
can reach higher densities than the dark matter (and is roughly
consistent with the relative densities in our galaxy \citep{binmer}).
We can then evaluate $n_0$ by using equation (\ref{rhoc}) to specify
$\rho_{\rm c}$, which depends on the fluctuation amplitude $Q$. The
scale length $r_{\rm dm}$ for the dark matter halo is then given by
equation (\ref{rdark}) and the corresponding scale for the baryonic
component is given by equation (\ref{rbary}).  The remaining parameter
that must be specified is the additional factor $f_{\rm vir}$ that
enhances the density (decreases the scale length). As reviewed in
\citep{tegrees}, $f_{\rm vir} \sim 1$ for large halo masses near 
$M = M_{\rm eq}$ and decreases to $f_{\rm vir} \sim 0.03$ for a
galactic mass halo $M \sim 2 \times 10^{12} M_\odot$ (where this
latter value is close to the mass of the Local Group \citep{penar}). 
Here we model this additional collapse factor with a simple scaling
law of the form $f_{\rm vir}=(M/M_{\rm eq})^{3/5}$.  Note that more
sophisticated models can be used, where the density scale is written
in terms of the halo concentration, which is then fit to numerical
simulations \citep{prada}.

\begin{figure}[tbp]
\centering 
\includegraphics[width=.90\textwidth,trim=0 150 0 150,clip]{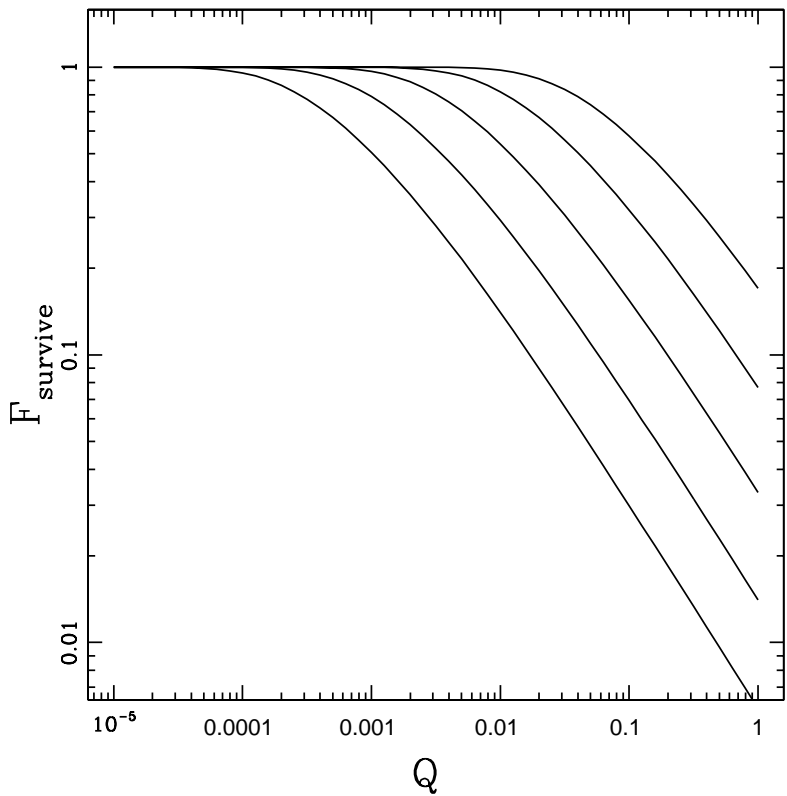}
\caption{\label{fig:survive} Fraction of solar systems that survive 
disruption from scattering encounters in galaxies of varying masses as
a function of the fluctuation amplitude $Q$. In universes with larger
$Q$, structure forms earlier, galactic structures are denser, and 
solar systems are more easily disrupted. The five curves shown
correspond to galaxy masses $M=10^{10}-10^{14} M_\odot$ (from lower
left to top right). }
\end{figure}

In a galaxy with a given total mass $M$, a fraction of the constituent
solar systems will reside at sufficiently large galactocentric radii
to survive disruption of their planet orbits. Figure \ref{fig:survive}
shows the resulting survival fractions for galaxies with masses in the
range $M=10^{10}-10^{14}M_\odot$ (from lower left to top right in the
figure).  The survival fraction is shown as a function of the
fluctuation amplitude $Q$. As the value of $Q$ increases, large scale
structure forms earlier and the resulting structures are denser
(characteristic of the background density of the universe at their
formation epoch).  Although solar systems are more easily disrupted
for galaxies in universes with large $Q$, some fraction of the systems
always survive. Note that we present results for density fluctuation
amplitudes up to $Q\to1$, but the formalism breaks down before
reaching this limiting value (see the discussion of Section
\ref{sec:discussion}; see also \citep{tegrees}). For completeness, 
we also note that the number of surviving solar systems in a galaxy 
of mass $M$ is given by 
\be
N_{\rm survive}= F_{\rm survive} \,\,
{\epsilon_{\rm sf} M / \langle M_\ast \rangle} \,,
\ee
where $\epsilon_{\rm sf}$ is the star formation efficiency and 
$\langle M_\ast \rangle$ is the mean stellar mass.  

The survival fraction depends on galactic mass. Within the framework
used here, galactic structures with smaller masses (e.g.,
$M=10^{10}M_\odot$, lower left curve) collapse to a greater degree,
have smaller values of $f_{\rm vir}$, and become more dense (see
equation [\ref{rhoc}]). Even in the most unfavorable case, with the
smallest galactic mass and the limiting value $Q \to 1$, the fraction
of surviving solar systems is $\sim0.003$, so that the galaxy is
expected to have $\sim3\times10^6$ viable systems. For a galaxy with a
mass comparable to that of the Milky Way, $M=10^{12} M_\odot$ (center
curve), the fluctuation amplitude must be larger than about
$Q\simgreat0.01$ in order for more than half of the solar systems to
be disrupted; in the limit $Q\to1$, about 3 percent of the systems
survive, corresponding to $\sim3\times10^9$ viable solar systems per
galaxy. 

\section{Radiation Constraints} 
\label{sec:radiation} 

In this section we find the radiation fields produced by the dense
galaxies under consideration and estimate the fraction of solar
systems that are disrupted through overheating. For this assessment,
we assume that the interstellar medium of the galaxy is optically thin
to its stellar radiation. This condition is expected to hold because
these dense galaxies are likely to experience efficient star
formation.  If enough gas and dust remain in the galaxy, however, the
interstellar medium could be optically thick, and the material would
become even hotter in order to transport out the luminosity of the
galaxy. This treatment thus represents a lower limit to the disruptive
influence of the background radiation. On the other hand, we are
implicitly assuming that the galaxy remains young enough that the
stars have not yet burned out \citep{dyson,al1997}. 

To start, we consider a rough estimate for the effects of radiation.
The total flux $F_G$ due to the background galaxy, at a particular
location, is given by the sum over all of the constitutent stars, 
\be
F_G = \sum_{\rm j=1}^N {L_j \over 4 \pi r_j^2} \,,
\label{sum} 
\ee 
where $r_j$ is the distance from the given location to the $jth$ star 
and $L_j$ is its luminosity.\footnote[2]{Here and throughout the paper, 
$F_G$ is the magnitude of the total flux passing through the top of the 
atmosphere, and is evaluated at the location of the planet. Note that 
the flux of radiation outward through the galaxy is much smaller.} 
To obtain an order of magnitude estimate, we first assume that the
stars are distributed spherically. The total number of stars $N$ is
then given by 
\be
N = n_\ast {4\pi \over 3} R^3 \,,
\ee
where $R$ is the size of the stellar system. We can replace 
the sum in equation (\ref{sum}) by an integral, i.e., 
\be
F_G = \int_0^R 4\pi r^2 dr n_\ast {\lbar \over 4 \pi r^2} = 
n_\ast \lbar R \,,
\ee
where $\lbar$ is the appropriate average value of the stellar
luminosity (see the discussion below). In order for the background
flux of the Galaxy to exceed that of the host star, we need 
\be
n_\ast \lbar R > {L \over 4\pi \varpi^2} \qquad {\rm or} \qquad 
4\pi \varpi^2 n_\ast R {\lbar \over L} > 1 \,,
\ee
where $\varpi$ is the orbital radius of the planet (assumed here to be
in a circular orbit).  We can eliminate $R$ in favor of the number $N$
of stars in the system,
\be
4\pi \varpi^2 n_\ast^{2/3} \left( {3 \over 4\pi} \right)^{1/3} 
N^{1/3} {\lbar \over L} > 1 \,. 
\ee
In the Galaxy today, this quantity is about $3 \times 10^{-6}$ (for
the choice $\lbar/L$ = 10; see below). The radiation flux from the
background galaxy thus becomes too intense for habitability when the
stellar density is increased by a factor of $\sim3\times10^5$ (note
that this density enhancement is roughly comparable to that needed for
scattering encounters to compromise planetary orbits, as found in
Section \ref{sec:scatter}).

\subsection{Radiation Fields within Galaxies} 
\label{sec:radfield} 

In this section we construct the expected radiation fields for
galaxies with extended structure, where we use the results from
Section \ref{sec:galstructure} to specify the density profiles. For a
solar system located at the center of the galaxy, the radiation field
from the background has the form 
\be
F_G = \int_0^\infty 4\pi r_0^3 \xi^2 d\xi {n_0 \over \xi (1+\xi)^3} 
{\lbar \over 4 \pi r_0^2 \xi^2} = n_0 r_0 \lbar
\int_0^\infty {d\xi \over \xi (1+\xi)^3} \,,
\ee
where the scale length $r_0=r_{\rm b}$ is that of the baryonic component. 
Note that the integral is logarithmically divergent. As a result, 
for a solar system exactly at the galactic center, one must include 
a cutoff radius. 

In practice, however, solar systems will not reside at the center.
For a solar system with dimensionless radial position $\xi=a$, the
expression for the background flux takes the form 
\be
F_G = {1 \over 2} n_0 r_0 \lbar
\int_0^\infty {\xi d\xi \over (1+\xi)^3} 
\int_{\rm -1}^1 {d\mu \over \xi^2 + a^2 - 2\xi a \mu} \,,
\ee
where $\mu = \cos\theta$ (and $\theta$ is the polar angle in 
spherical coordinates). The angular integral can be evaluated
to obtain 
\be
F_G = {1 \over 4a} n_0 r_0 \lbar
\int_0^\infty {d\xi \over (1+\xi)^3} \left\{ 
\log \left[ \xi^2 + a^2 + 2\xi a \right] - 
\log \left[ \xi^2 + a^2 - 2\xi a \right] 
\right\} \,.
\ee
The integral can be evaluated so the flux has the form
\be
F_G = {n_0 r_0 \lbar \over 2(a^2-1)} 
\left[ 1 - {2 \log a \over a^2-1} \right] \,.
\label{galflux} 
\ee

\begin{figure}[tbp]
\centering 
\includegraphics[width=.90\textwidth,trim=0 150 0 150,clip]{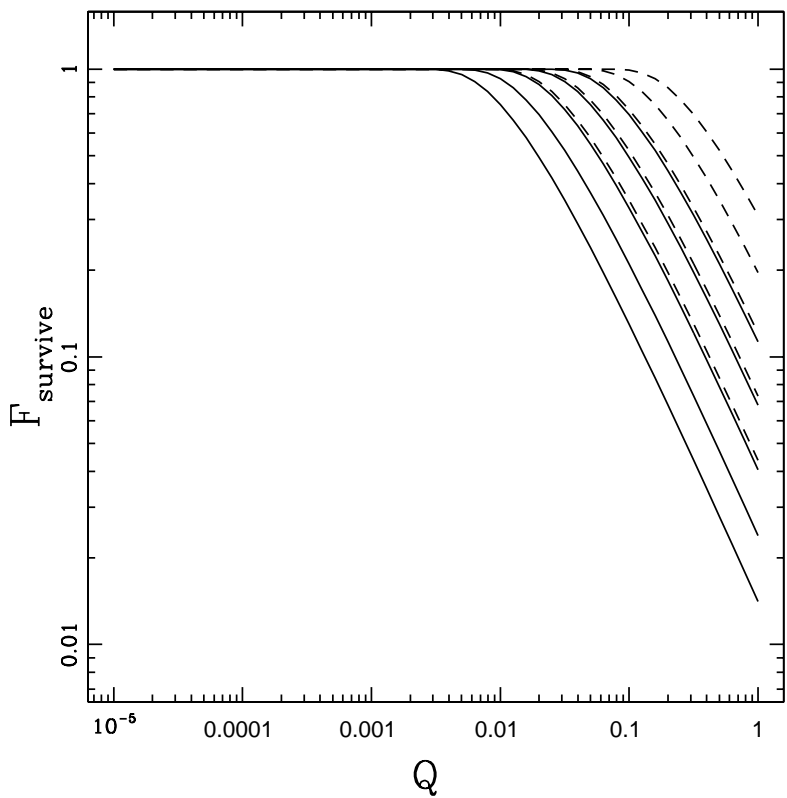}
\caption{\label{fig:survrad} Fraction of solar systems that survive
disruption from radiation in galaxies of varying masses as a function
of the primordial fluctuation amplitude $Q$.  As the value of $Q$
increases, structure forms earlier, so that galactic structures are
denser and galactic background radiation is more intense. Results are
shown for $\lbar/L$ = 10 (solid curves) and 1 (dashed curves). For 
each case, the five curves correspond to (total) galaxy masses $M$ 
= $10^{10}-10^{14}$ $M_\odot$ (from lower left to top right). } 
\end{figure}

\subsection{Survival Fractions for Radiation} 
\label{sec:radsurvive}  

Next we use the results outlined above to estimate the fraction of
solar systems that will reside in regions of the galaxies where the
background radiation field does not compromise habitability.  The
requirement that the radiation from the host star, at the location of 
the planet, exceeds that of the background galaxy takes the form  
\be
(a^2-1) \left[ 1 - {2 \log a \over a^2-1} \right]^{-1} 
> n_0 r_0  2\pi \varpi^2 {\lbar \over L} \,, 
\label{radcon} 
\ee
where $a$ is the dimensionless radial location within the galaxy. 

To evaluate this constraint, we need to specify the quantities on the
right hand side of equation (\ref{radcon}).  For this estimate, the
number density scale $n_0$ and length scale $r_0$ were determined as
described in Section \ref{sec:scatter}, where we considered survival
due to planetary scattering. We also need to specify the relative
brightness of the stellar population through the ratio $\lbar/L$,
which measures the power output of the average star. In our present-day
universe, the mass to light ratios for typical galaxies are of order
$1-10$ in solar units \citep{binmer,bintrem}, where this accounting
includes all contributions to the mass (i.e., dark matter, gas, and
stellar remnants in addition to main-sequence stars). We thus expect
the ratio $\lbar/L$ to be of order unity, or perhaps somewhat larger.
Here we present results for $\lbar/L$ = 1 and 10. Note that the
corresponding light to mass ratio for zero-age main-sequence stars
would imply a much larger ratio $\lbar/L\sim1000$ \citep{leitherer};
the galactic average value, appropriate in this context, is lower
because the high mass stars die out quickly and stellar remnants are
included. We also note that the era of nuclear burning stars lasts for
trillions of years \citep{al1997}, but not forever, so that the ratio
$\lbar/L$ will eventually become much smaller.

Figure \ref{fig:survrad} shows the resulting fraction of solar systems
that survive the intense radiation fields of their host galaxies,
shown here as a function of the amplitude $Q$ of the primordial
density fluctuations.  These results are qualitatively similar to
those obtained previously for disruption due to scattering of
planetary orbits (compare with Figure \ref{fig:survive}). However,
scattering interactions are somewhat more disruptive, so that the
survival fractions shown in Figure \ref{fig:survrad} (considering the
radiation fields) are higher. The most extreme case shown corresponds
to the smallest galaxy with mass $M=10^{10}$ $M_\odot$, the larger
value of the luminosity ratio $\lbar/L$, and the limiting value of the
fluctuation amplitude $Q\to1$; even in this case, however, 1\% of the
solar systems are projected to survive the disruptive effects of
radiation. 

\subsection{Allowed Regions of Galactic Parameter Space} 
\label{sec:pspace} 

The previous sections outline the requirements necessary for a solar
system to survive disruption due to scattering by background stars
(Section \ref{sec:survive}) and intense radiation fields (Section
\ref{sec:radsurvive}).  These results show the fraction of surviving
solar systems as a function of the fluctuation amplitude $Q$ (see
Figures \ref{fig:survive} and \ref{fig:survrad}). In this section, we
delineate the regions within a galaxy that allow solar systems to
survive these two channels of disruption.

Survival from scattering requires that the scattering optical depth
$\tau_S$ is less than unity, 
\be
\tau_S = {n_0 \cross v t_{\rm c} \over \xi (1+\xi)^3} < 1\,.
\label{tauscatter} 
\ee
Survival from radiation requires that the radiation received from 
the background galaxy is less than that of the host star, i.e.,  
\be
{\cal R} = {F_G \over F_\ast} = {n_0 r_0 2\pi \varpi^2 \over \xi^2-1} 
{\lbar \over L} \left[1-{2 \log\xi\over\xi^2-1}\right] < 1\,. 
\label{radsurvive} 
\ee
Because planetary orbits can have different semimajor axes, this
condition must be considered as approximate. Moreover, as considered
in Section \ref{sec:ghz}, the background galaxy could provide all of
the power for a habitable planet. The constraint of equation
(\ref{radsurvive}) thus applies to conventional habitability.

\begin{figure}[tbp]
\centering 
\includegraphics[width=.90\textwidth,trim=0 150 0 150,clip]{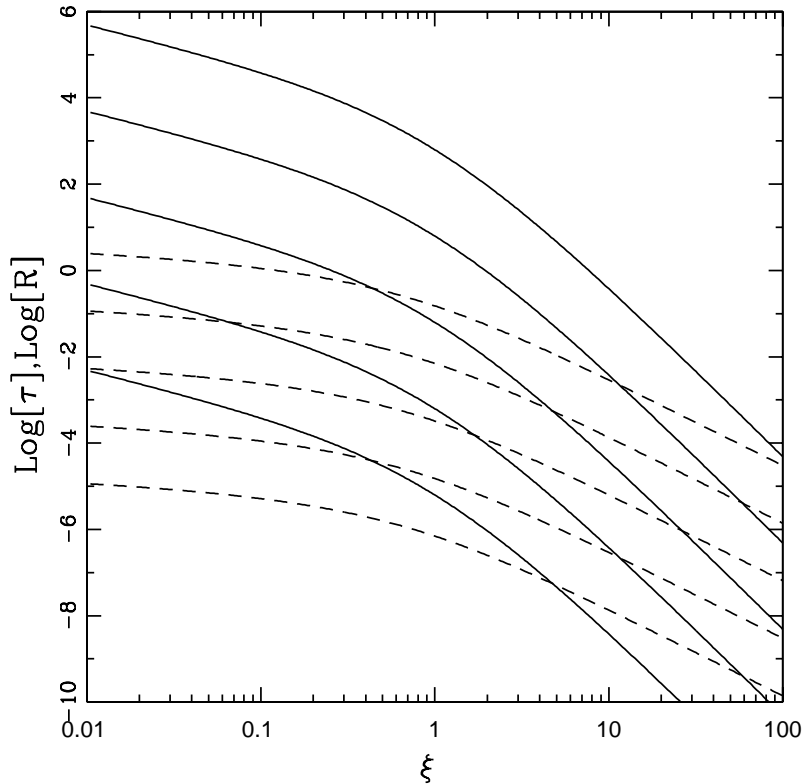}
\caption{\label{fig:ratios} Disruption levels as a function of position 
within an idealized galaxy. The solid curves show the probablity of
scattering an Earth-like planet (over $t_{\rm c}$ = 1 Gyr) for galaxy models
with varying density scales $n_0$. The dashed curves show the ratio of 
the radiation flux from the background galaxy to that provided by a 
central sun-like star. For both cases, curves are shown for galaxies
with density scales $n_0$ = 1 (bottom), $10^2$, $10^4$, $10^6$, and 
$10^8$ (top), in units of pc$^{-3}$. }
\end{figure} 

Both the scattering optical depth $\tau_S$ and the radiation ratio
$\cal{R}$ are functions of radial position $\xi$ within the galaxy.
These functions are shown in Figure \ref{fig:ratios} for varying
choices of the density scale $n_0$. The lower curves in Figure
\ref{fig:ratios} correspond to values comparable to those of the solar
neighborhood in the galaxy today.  The remaining curves correspond to
denser galactic environments, where the stellar density parameter is
increased by factors up to $10^8$.

A number of trends are clear from Figure \ref{fig:ratios}. First, for
density enhancements of order $10^6$, the range of allowed parameter
space starts to be compromised.  Second, disruption by scattering
interactions is usually more important than disruption by radiation.
However, in the outer parts of the galaxy, the radiation fields
provide the dominant channel of disruption. This behavior arises
because the radiation flux decreases with radius as $1/\xi^2$ whereas
scattering disruption varies as $1/\xi^4$ (where both results are
evaluated in the limit $\xi\gg1$).  Finally, we note that the outer
parts of the galaxy are always diffuse enough to allow survival in the
face of both scattering and radiation, i.e., some fraction of the
solar systems always survive.

Within the context of the formalism used here, the outskirts of the 
galaxy allow survival. In practice, however, the galaxy does not 
continue outward forever, but rather must have a most distant star. 
The fraction of stars contained within the dimensionless radius $\xi$  
is given by 
\be
{N(\xi) \over N} = {\xi^2 \over (1+\xi)^2} = 1 - 
{1 + 2\xi \over (1+\xi)^2} \,, 
\ee
where $N$ is the total number of stars.  In order of magnitude, the
outermost star in the galaxy resides at a radial location $\xi$ given
by $N(\xi)=N-1$, or equivalently, 
\be
{1 + 2\xi \over (1+\xi)^2} = {1 \over N} \,, 
\ee
which has the solution
\be
\xi = N - 1 + \sqrt{(N-1)^2+N-1} \approx 2N \,. 
\ee
In order for the galaxy to be so dense that even the outermost
star experiences disruption through solar system scattering, 
the requirement becomes 
\be
n_0 \cross v t_{\rm c} > (2N) \left[ 1 + 2N \right]^3 \approx (2N)^4\,. 
\ee 
For a typical galaxy, the right hand side of this expression is
$\sim10^{44}$, so that the density scale $n_0$ must be larger than
that of our galaxy by a factor of $\sim10^{40}$ in order for no solar
systems to survive. Using simple scaling arguments, the density
$n_0{\propto}Q^3$, so that the largest enhancement factor is of order
$10^{15}$ ($\ll 10^{40}$), so that galaxies are never dense enough to 
render all of their solar systems uninhabitable. 

Note that the preceding calculation assumes that the stellar component
of the galaxy does not have a well-defined outer edge.  In practice,
however, the outer parts of the galaxy can become too rarified to form
stars \cite{schaye}, where the intergalactic radiation field prevents
the formation of molecules.  As a result, the outermost portions of
the galaxy would have to be populated with stars via dynamical relaxation.

Before leaving this section, we note that scattering and radiation are
roughly comparable as disruptive influences on habitable solar systems
for galaxies in the present-day universe.  We can understand this
apparent coincidence as follows: The scattering optical depth is given
by the product $n_0\sigma{v}t_{\rm c}$ (from equation
[\ref{tauscatter}]), whereas the analogous product that determines
radiative disruption is $\sim{n_0}r_0\pi\varpi^2$ (from equation
[\ref{radsurvive}]). Both of these quantities thus depend linearly on
the stellar density scale $n_0$. In addition, the cross section
$\sigma$ for scattering is roughly comparable to the area of the
planetary orbit $\pi\varpi^2$, which determines how much radiation is
received from the central star. The coincidence is that the total
distance traveled by a solar system during the benchmark time scale
($vt_{\rm c}$) is comparable to the distance scale $r_0$ of the
galactic density profile (i.e., the orbit time $r_0/v\sim t_{\rm c}$).
More compact galaxies --- and especially their central regions --- 
have shorter orbit times and are more disruptive via scattering 
(compared to radiation). 

\section{Galactic Habitable Zones} 
\label{sec:ghz} 

The intense radiation fields provided by compact galaxies introduce
another possible channel for habitable planets. If the galactic
background radiation has the proper intensity, then any potentially
habitable planet can have the right temperature to support liquid
water on its surface (one of the usual requirements for habitability
\citep{kasting}). More specifically, the surface temperature of a
planet will have a minimum value, independent of its orbit within its
host solar system (including unbound orbits). As expected, and as
shown in Section \ref{sec:radiation}, the background radiation flux
within a galaxy decreases with galactocentric distance, so that a
spherical shell within the galaxy will provide the proper radiations
levels. We denote this region as the Galactic Habitable Zone (GHZ).

\begin{figure}[tbp]
\centering 
\includegraphics[width=.90\textwidth,trim=0 150 0 100,clip]{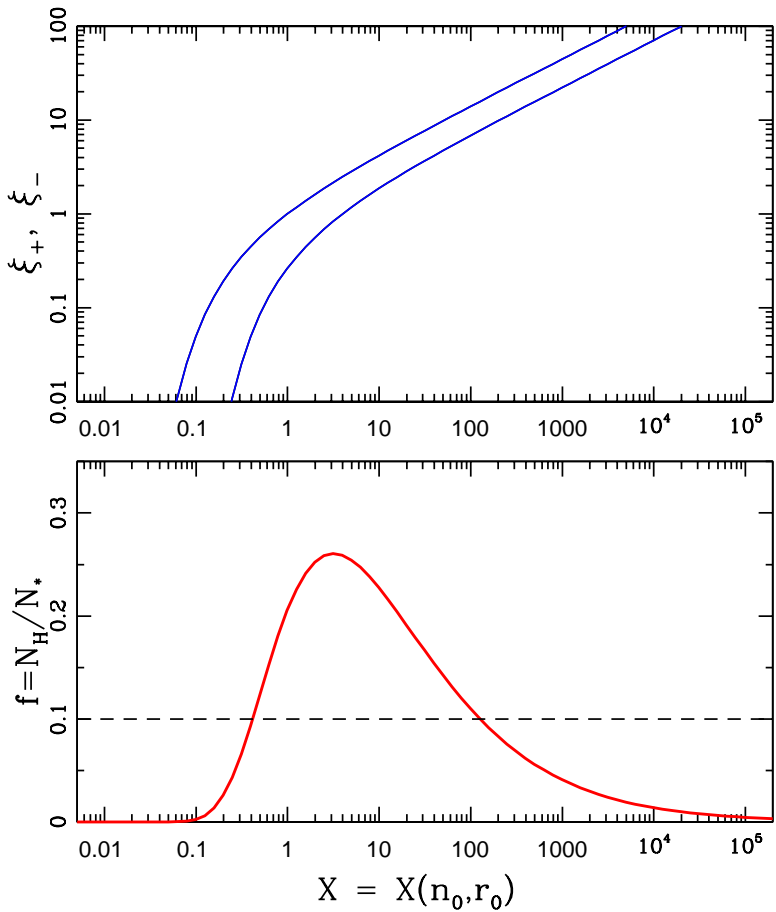}
\caption{\label{fig:ghz} The Galactic Habitable Zone is the portion 
of a galaxy where the background radiation has the right intensity to 
support biospheres, independent of the orbital elements of the planet.
The top panel shows the inner and outer boundaries of the GHZ as a
function of the composite parameter $X$ that encapsulates the relevant
galactic properites (see text). The bottom panel shows the fraction
$f$ of the solar systems (per galaxy) that reside in the GHZ. If each
star produces on average 10 planets, then galaxies with $f>0.1$ (above 
the dashed line) will support more potentially habitable planets than
galaxies in our universe. }   
\end{figure} 

\subsection{Extent of the Galactic Habitable Zone} 
\label{sec:extent} 

We start by delineating the boundaries of the GHZ.  Section
\ref{sec:radiation} shows that the background galaxy provides a
radiative flux $F_G$ given by equation (\ref{galflux}). By equating 
this flux to that required for habitability, we obtain a constraint 
of the form 
\be
F_G = C {L_\odot \over 4 \pi \varpi^2} \,,
\ee
where $\varpi$ is the semimajor axis of the Earth-analog orbit and the
factor $C$ is a dimensionless constant of order unity. The requirement
of liquid water on the surface of a planet depends on the planetary
atmosphere and other properties. Here we use the requirement that the
background flux of the galaxy $F_G$ is within a factor of two of that
provided by the Sun to the Earth, i.e., we allow $C$ to vary over the
range $0.5\le{C}\le2$. Next we define a composite parameter $X$ that 
incorporates the galactic properties, 
\be
X \equiv 2 \pi \varpi^2 n_0 r_0 {\lbar \over L_\odot} \approx 
1.57 \left( {n_0 \over 10^6 \, {\rm pc}^{-3}} \right) 
\left( {r_0 \over 1 \, {\rm kpc}} \right) 
\left( {\lbar \over 10 L_\odot} \right) \,. 
\ee
The Galactic Habitable Zone then lies between radial locations 
$\xi=\xi_{\rm \pm}$ that statify the requirement
\be
{1 \over \xi^2 - 1} \left[ 1 - {2\log\xi \over \xi^2-1} \right]
= {C \over X}\,,
\ee
where $C$ = 1/2 for the outer boundary and $C$ = 2 for the 
inner boundary of the GHZ. 

The resulting boundaries of the GHZ are shown in the upper panel of
Figure \ref{fig:ghz} as a function of the composite parameter $X$.
The corresponding fraction $f=N_H/N_\ast$ of solar systems that reside
in the GHZ are shown in the lower panel of the figure. The horizontal
dashed line delineates the portion of parameter space for which 10
percent of the solar systems reside in the GHZ, which corresponds to
values of the parameter $X$ in the range $X\approx0.4-125$. Note that
nearly all of the planets in the GHZ are potentially habitable, by
definition. In contrast, within ordinary galaxies (like our own), only
a fraction of planets have orbits at the right distance from their
host stars to support biospheres. Moreover, this fraction is only
about 10 percent: The observed orbital spacing in solar systems is
such that only about one star per solar system has the right
temperature, and solar systems can often support $\sim10$ planets.
This result raises an interesting possibility: Galaxies with composite
parameter $X=1-100$ could support more habitable planets than
those in our universe.

The condition of having a robust GHZ requires the parameter $X$ to lie
in the range 1 -- 100, which requires that the column density of stars
in the galaxy $n_0r_0\approx10^8-10^{10}$ pc$^{-2}$ (where we take
$\lbar/L=1$).  For comparison, typical column densities for galaxies
in our universe are of order $n_0r_0\sim100$ (e.g., for a galaxy with
$N=10^{11}$ stars and scale length $r_0$ = 12.5 kpc). As another
benchmark, the surface brightness of our Galaxy is $\sim1000L_\odot$
pc$^{-2}$ \citep{binmer}, which converts to a column density in stars
of order 100 -- 1000 pc$^{-2}$. As a result, within the context of
this formalism, the most favorable galaxies for potentially habitable
planets must have column densities enhanced (relative to our universe)
by factors of $10^6-10^8$.

We would like to find the range of fluctuation amplitudes $Q$ that
produces galaxies with large GHZs. Unfortunately, there is not a
definitive one-to-one mapping between the constraints on galactic
properties and analogous constraints on the density fluctuation
parameter $Q$: Universes with a given $Q$ produce galaxies over a
range of mass scales, and different galactic masses collapse to a
range of densities and sizes. Nonetheless, we can make a rough
estimate using typical values for the galactic masses.  Since we
expect the density to scale with fluctuation amplitude according to
$n_0 \propto Q^3$ and the length parameter to scale according to 
$r_0 \propto 1/Q$, the column density scales as $Q^2$. The preferred
fluctuation amplitude $Q_{\rm opt}$ for habitable planets is thus
larger than that of our universe by a factor of $10^3-10^{4}$, so 
that the optimum amplitude is expected to fall in the range 
\be 
Q_{\rm opt} = 0.01 - 0.1 \,.  
\label{bestq} 
\ee 

The above scaling argument indicates that a substantial GHZ requires
relatively large values of the amplitude $Q$. In such universes,
however, several complications should be kept in mind: Large
fluctuation amplitudes can lead to rapid structure formation, and
could increase the nonlinear mass scale beyond that realized in our
Universe. If the component of dark energy --- which acts to accelerate
the universe --- is sufficiently small, then galaxies can readily
merge with each other, and the first generation of galaxies would not
survive long enough to support life. The question of habitability
would then depend on the nature of the resulting merged galactic
structures. In addition, the central regions of these dense galaxies
are likely to support the formation of massive black holes, which
would alter the galactic luminosity.  

Some authors have suggested that the probability of an inflationary
universe producing a given amplitude for the density fluctuations 
is an increasing function of $Q$ \citep{garriga}. This result is 
consistent with the finding that small values of $Q$ require carefully
chosen small parameters in the inflationary potential
\citep{bardeen,afg}. As a result, universes with larger values of
$Q$, including the range indicated in equation (\ref{bestq}), could 
be more common than our own.

\subsection{Properties and Consequences} 
\label{sec:consequence} 

For the larger stellar densities expected within the GHZ, planets have
a good chance of being scattered out of their orbits. Over the range
of galactic masses and the amplitudes $Q_{\rm opt}$ given in equation
(\ref{bestq}), anywhere from 3\% to 97\% of the planets in Earth-like
orbits will experience disruption (see Figure \ref{fig:survive}).
Planets on wider orbits will be disrupted to an even greater extent.
As a result, the majority of the viable planet population in this
scenario will be freely floating, rather than bound in orbit about
their host stars. Although scattering encounters are expected to
liberate planets in the GHZ, stellar collisions will be rare for time
scales $t_{\rm c}$.  At the high densities of the galactic centers,
however, stellar collisions can more readily take place. The centers 
of these galaxies are thus favorable sites for the formation of large
black holes \citep{bhrees}.

Freely floating planets, and entire solar systems, can eventually
leave the GHZ through accumulated perturbations from passing
stars. The time required for this process of dynamical relaxation is
given by
\be
t_{\rm relax} = {v^3 \over 8\pi\Lambda G^2 M_\ast^2 n_\ast} \,,
\label{trelax} 
\ee
where $v$ is the orbit speed and $\Lambda\equiv\ln[b_2/b_1]$, where
$b_1$ and $b_2$ are the minimum and maximum possible impact parameters
\citep{bintrem}. Using equation (\ref{trelax}) to evaluate the
relaxation time for galaxies with $X=1-100$ and for locations within
the GHZ, we find 
\be
t_{\rm relax} \sim 10^{5} \,\,{\rm Gyr}\,\,
\left({N_\ast \over 10^{12}}\right)^{1/2} 
\left({r_0 \over 0.1\,{\rm kpc}}\right)^{3/2} \xi^{5/2} \,,
\ee
where $N_\ast$ is the total number of stars in the galaxy. Since the
dimensionless radius $\xi\sim1$ in the GHZ, the relaxation time is
much longer than the characteristic time scale $t_{\rm c}$ = 1 Gyr
required for habitability. As a result, most potentially habitable
planets will stay within the GHZ long enough for biological evolution
to take place.\footnote[2]{Note that the central regions of the  
galaxies, where $\xi\ll1$, have much shorter relaxation times, 
so that rearrangement of stellar orbits is expected there.} 
 
In a related but different context, the concept of a Galactic
Habitable Zone has been put forth for our Galaxy \citep{gzone}. In
that case, however, the conditions required to compromise habitability
are much less extreme than those of this paper.  Specifically, the
habitable zone from \citep{gzone} is based on the chemical makeup of
the galaxy as a function of position (and requires the metallicity to
be within a factor of $\sim2$ of the value for our Solar System).
Since the metallicity $Z$ decreases with galactocentric distance, the
outer parts of the galaxy have low $Z$ and are less suited for life.
The definition of the GHZ used in this paper is similar in spirit to
the constraints of \citep{thompson}, who considered the disruption of
planetary systems in rich star clusters.  In this latter study,
sufficiently rich stellar environments produce intense radiation
fields that heat planet-forming disks above the temperature required
for ices to condense; with no ices, the process of giant planet
formation can be compromised.

This present work focuses on the survival of planetary systems after
they are made; future studies should also consider how environmental
effects can hinder their formation. In particular, the background
temperature within the GHZ allows for water to be in liquid form at
atmospheric pressures, which implies that the water could be gaseous
at the lower pressures characteristic of planet-forming disks
\citep{hayashi}. In this case, the dust grains -- and later the rocks
-- that provide the building blocks for planets will have little
water, and terrestrial planets could end up dry. Note that this
dilemma is a scaled up version of the same issue facing potentially
habitable planets in our universe: In the inner solar system, where
planets are warm enough to support biospheres, the rocky planets must
be built out of raw material with relatively little water. As a
result, alternate channels for the delivery of water to Earth, and
hence other terrestrial planets, have been considered. If enough of
the nebula remains intact after a terrestrial planet is formed, its
atmosphere can be both massive and rich in hydrogen. In this setting,
the atmospheric hydrogen can react with various oxides (such as FeO)
in the still-liquid rocky planet and produce a significant amount of
water \citep{sasaki,ikoma}. In any case, the delivery of water to
potentially habitable planets in the GHZ should be explored further.

For completeness, we note that the radiation fields found in the GHZ
are much less intense than those required to make the night sky as
bright as a stellar surface, the condition considered in Olbers'
paradox \citep{harrison,wesson}. In order for any given line of sight
to intercept a stellar surface, the optical depth through the galaxy
would have to exceed unity, so that $n_0r_0{\pi}R_\ast^2\sim1$
(compare with the analogous expressions for disruption from equations
[\ref{tauscatter}] and [\ref{radsurvive}]).  However, stellar radii
$R_\ast$ are much smaller than planetary orbits ($\varpi\sim1$ AU) or
scattering cross section length scales ($\ell\sim\sqrt{\sigma}\sim1$
AU). As a result, the column density $n_0r_0$ of stars through the
galaxy would have to be much larger than the values considered here,
by a factor of $(\varpi/R_\ast)^2 \sim 4\times10^4$, for the night sky
to be as bright as the surface of the Sun.\footnote[2]{In other words,  
in the GHZ, the night sky is (approximately) as bright as the daytime
sky on Earth, but not as bright as a stellar photosphere.}  This
distinction is important: The incoming radiation has the energy
characteristic of stellar photospheres (with temperatures
$T_\ast\approx3000-10,000$ K), whereas the planets have much lower
temperatures $T_P\approx273-373$ K (and emit lower energy, infrared
radiation).  This temperature difference allows for a heat engine to
operate and thereby perform the work necessary to run a biosphere.

In order for a biosphere to operate, the heat generated by the planet
must be able to radiate away. In our present-day universe, the
background temperature of the galaxy ($T_{\rm gal}\sim3-10$ K) is much
colder than the planetary temperature $T_P$, so that exhaust heat
poses no problem. Since the background temperature of the galaxy is
hotter in the GHZ, one might worry that a heat engine could have
trouble operating.  As outlined above, however, the galaxy is
optically thin to its internal radiation, including both stellar
radiation and that re-radiated by planets. As a result, the ultimate
heat sink for the galaxy is cold extragalactic space (where
$T\ll{T_P}$).  Planets can thus absorb optical light and freely
re-radiate the energy as infrared radiation.  This process generates
large amounts of entropy, which allows for a decrease in the entropy
in parts of the biosphere itself.  In addition, planetary atmospheres
raise the temperature of the planets further, and this increase also
helps to radiate the necessary exhaust heat. For completeness, we also
note that a biosphere does not function as a simple carnot heat
engine, so that its efficiency is not given by the classical
expression ($\epsilon=1-T_c/T_h$). Instead, one must consider the
Gibbs free energy of the system, including all of the terms associated
with the different particle components and their chemical potentials
(for further discussion, see \citep{lunine,scharf}).

\section{Conclusion} 
\label{sec:conclude} 

Extending previous work, this paper has shown that when galactic
stellar systems become sufficiently dense relative to those in our
present-day universe, significant disruption of solar systems can take
place. Such disruption can render Earth-like planets uninhabitable
through two different mechanisms: Scattering interactions can perturb
planetary orbits before life has time to evolve and radiation from the
background galaxy can dominate that of the host star and thereby make
all planets too hot to support biospheres. Nonetheless, some fraction
of the potentially habitable systems will survive.  A summary of our
results is given below (in Section \ref{sec:summary}), followed by a
discussion of their implications and indications for further work
(Section \ref{sec:discussion}).

\subsection{Summary of Results} 
\label{sec:summary}   

We have performed a large ensemble of $\sim10^6$ numerical simulations
to determine the interaction cross sections for the disruption of
planetary orbits via encounters with passing stars (see Figure
\ref{fig:csection}). The disruption cross section of Earth-analog 
orbits can be fit with the function given by equation
(\ref{crossfit}). As expected, this cross section approaches the
asymptotic form $\cross\sim1/v^2$ in the high speed limit, where the
limiting regime requires $v \gg 1000$ km s$^{-1}$.

The density of stars within a galaxy is a function of position; here
we use the Hernquist profile of equation (\ref{hernquist}) as a
working model. For this class of profiles, both the scattering rate
$\Gamma$ and the background radiation flux $F_G$ diverge in the inner
limit $\xi\to0$. On the other hand, both $\Gamma\to0$ and $F_G\to0$ in
the limit of large radii $\xi\gg1$.  Inclusion of galactic structure
thus changes the problem from previous treatments, which implicitly
assume that galaxies are characterized by a single density.  In every
galaxy, some fraction of the solar systems will be disrupted by
scattering encounters, whereas some other fraction will be disrupted
by background radiation fields (and some fraction will survive). 

The characteristic densities for galactic structure (see equation
[\ref{rhoc}]) increase with the amplitude $Q$ of the primordial
density fluctuations. As a result, the fraction of solar systems
within a galaxy that survive disruption -- and thereby remain
potentially habitable -- is a decreasing function of $Q$. This trend
is shown in Figure \ref{fig:survive} for disruption of planetary
orbits by scattering encounters and in Figure \ref{fig:survrad} for
disruption by the background radiation field of the galaxy. Note that
the survival fractions also depend on galactic mass (as shown in the
figures). The outer parts of the galaxies remain habitable for
essentially any choice of the galactic structure parameters (see
Figure \ref{fig:ratios}).

Because of the intense galactic background radiation fields, some
portion of the galaxy can provide the right flux levels to support
habitable planets. Any planets residing in this region -- denoted here
as the Galactic Habitable Zone -- can have surface temperatures
compatible with maintaining liquid water. Although planets residing
too close to their host stars will still be too hot, planets in all
other orbits, including freely floating bodies, can in principle
support biospheres.  Moreover, the GHZ is large enough so that some
galaxies in other universes can support more potentially habitable
planets (powered by the background radiation of the galaxy) than the
galaxies in our present-day universe (Figure \ref{fig:ghz}).

\subsection{Discussion} 
\label{sec:discussion} 

In approximate terms, disruption of habitable planets through both
scattering and radiation becomes important when the background density
of stars is a million times larger than that of the Solar
Neighborhood, which is not atypical within galaxies of our present-day
universe. Since galactic densities increase with the approximate
scaling $\rho \propto Q^3$ (equation [\ref{rhoc}]), we naively expect
the disruption of habitable planets to become important when $Q$ is
about 100 times larger than in our universe, i.e., for $Q \simgreat
10^{-3}$. This expectation is borne out in Figures \ref{fig:survive}
and \ref{fig:survrad}, although universes can remain habitable for
even larger values of $Q$ if we only require that a fraction of the
solar systems remain viable.

The results of this work indicate that habitable planets can survive
in galaxies with a wide range of properties, and hence in a wide range
of universes. At first glance, this result may seem to contradict
previous work that places relatively tight limits on the fluctuation
amplitude $Q$. In this current generalized treatment, however, more
planets can survive for several reasons: [A] Galactic structures
naturally include a wide range of stellar densities, with lower
densities in the outer regimes that allow for habitable orbits to
survive. [B] With the consideration of internal galactic structure, we
can determine the fraction of solar systems that avoid disruption
(rather than using an all-or-nothing approach). Because planets are
thought to be common, and galaxies contain many solar systems, a
galaxy (universe) can remain habitable with only a small fraction of
its solar systems surviving (note that Figures \ref{fig:survive} and
\ref{fig:survrad} are plotted with logarithmic scales). [C] Even
planets that are stripped from their host stars can remain habitable
if they reside within the Galactic Habitable Zone, i.e., the region of
the galaxy where the galactic background radiation has the proper
intensity to heat planets.

The results of this paper are presented for density fluctuation
amplitudes as large as $Q=1$.  However, the reader should keep in mind
that the formalism breaks down before reaching such high values. For
sufficiently large fluctuation amplitudes, structure formation is
thought to take place violently, thereby leading to rampant black hole
formation and other departures from the standard scenario.  This
regime of structure formation has not been well-studied via numerical
simulations, but such work should be carried out in the future. Even
for the regime of moderately higher values of the amplitude $Q$,
numerical simulations of structure formation should be carried out in
order to verify the assumption that dark matter halos approach a
nearly universal form (assumed here to be NFW/Hernquist). We also need
a better specification of what happens to the baryons in these
high-$Q$ scenarios. In this work, we assume that galaxies become
scaled-up versions of the galactic bulges in our universe (with
density profiles of equation [\ref{hernquist}]). If the galactic
system has enough angular momentum, however, baryons can settle into a
rotationally supported disk structure (like the spiral galaxies of our
universe). The survival fractions and background radiation fields
should also be determined for galaxies with disk geometries.

This treatment considers universes with different amplitudes $Q$ for
the density fluctuations, and hence denser galactic structures, but
holds fixed the other properties of the universe. In general, however,
both the fundamental constants of physics and cosmological variables
(in addition to $Q$) could be different in other universes.  Such
additional variations would change the predicted fraction of habitable
solar systems. As one example, if the gravitational constant $G$ is
larger, then stars are brighter \citep{chandra,phillips,adams}, and
habitable planets must have larger orbits; as a result, planetary
scattering could be enhanced. On the other hand, this change could be
offset with an increase in the fine structure constant $\alpha$, which
would act to suppress nuclear fusion and would require closer orbits
to achieve chemical reactions \citep{bartip,burost}. Future work
should thus take into account coupled variations in all of the
relevant parameters.  In addition, the `constants' of nature could
also vary with time within a particular universe \citep{barrow,uzan}.
Considering all of the possibilities is an enormous undertaking; the
first step is to consider more limited possible variations as carried
out here.

Finally, we note that most considerations of alternate universes use
the properties of our own universe as the baseline for comparison,
often with an implicit assumption that our universe provides the best
possible environment for the development of life. The results of this
paper call this assumption into question: With the right parameters,
galaxies can support extensive Galactic Habitable Zones, where up to
20\% of the stars, and hence up to 20\% of all planets, reside within
a radiation field comparable to the one the Earth receives from our
Sun. Such galaxies -- residing in universes with the right properties
-- could support more habitable planets than our own. Within our
universe, Earth is often considered as the optimal planet for
supporting a biosphere, but other planets could be even more favorable
\citep{heller}. This issue, which could be called superhabitability,
should also be explored further.

\acknowledgments

We would like to thank Konstantin Batygin, Juliette Becker, Gus
Evrard, Evan Grohs, Gordy Kane, Minhyun Kay, Jake Ketchum, and Chris
Spalding for useful discussions and suggestions, as well as two 
anonymous referees.  This work was supported by JT Foundation grant
ID55112 ``Astrophysical structures in other universes'' and by the
University of Michigan. FCA is also supported by FQXi grant MGB-1414;
AMB is also supported by NSF grants INSPIRE-1363720 and DMS-1207693,
and by the Simons Foundation.


\begin{thebibliography}{99} 

\bibitem{cobe} 
G. F. Smoot et al., 
{\it Structure in the COBE Differential Microwave Radiometer First-year Maps}, 
{\sl Astrophys. J.} {\bf 396} (1992) L1

\bibitem{wmap} 
D. N. Spergel et al., {\it Three-Year Wilkinson Microwave Anisotropy}
{\it Probe (WMAP) Observations: Implications for Cosmology},
{\sl Astrophys. J. Suppl.} {\bf 170} (2007) 377 

\bibitem{planck}
Planck Collaboration: P.A.R. Abe et al.,
{\it Planck 2013 Results. XVI: Cosmological parameters}, 
{\sl Astron. Astrophys.} {\bf 571} ( 2014) 16

\bibitem{guth} 
A. H. Guth, {\it Inflation and Eternal Inflation}, 
{\sl Phys. Rep.} {\bf 333} (2000) 555  

\bibitem{bardeen} 
J. M. Bardeen, P. J. Steinhardt, and M. S. Turner, 
{\it Spontaneous Creation of Almost Scale-free Density Perturbations}
{\it in an Inflationary Universe}, {\sl Phys. Rev.} D {\bf 28} (1983) 679 

\bibitem{afg}
F. C. Adams, K. Freese, and A. H. Guth,
{\it Constraints on the Scalar-field Potential in Inflationary Models},  
{\sl Phys. Rev.} D {\bf 43} (1991) 965 

\bibitem{garriga} 
J. Garriga and A. Vilenkin, 
{\it Anthropic Prediction for $\Lambda$ and the $Q$ Catastrophe}, 
{\sl Prog. Theor. Phys.} {\bf 163} (2006) 245 

\bibitem{tegrees} 
M. Tegmark and M. J. Rees, 
{\it Why Is the Cosmic Microwave Background Fluctuation Level $10^{-5}$?} 
{\sl Astrophys. J.} {\bf 499} (1998) 526   

\bibitem{tegmark}
M. Tegmark, A. Aguirre, M. J. Rees, and F. Wilczek, 
{\it Dimensionless Constants, Cosmology, and other Dark Matters}, 
{\sl Phys. Rev.} D {\bf 73} (2006) 3505

\bibitem{reesost} 
M. J. Rees and J. P. Ostriker, 
{\it Cooling, Dynamics and Fragmentation of Massive Gas Clouds:}
{\it Clues to the masses and radii of galaxies and clusters},  
{\sl Mon. Not. R. Astron. Soc.} {\bf 179} (1977) 541  

\bibitem{whiterees} 
S.D.M. White and M. J. Rees, {\it Core Condensation in Heavy Halos:}
{\it A two-stage theory for galaxy formation and clustering},  
{\sl Mon. Not. R. Astron. Soc.} {\bf 183} (1978) 341 

\bibitem{rees}
M. J. Rees, {\sl Before the Beginning}, Perseus (1997) 

\bibitem{reessix}
M. J. Rees, {\sl Just Six Numbers}, Basic Books (2000)

\bibitem{vilenkin} 
A. Vilenkin, 
{\it Unambiguous Probabilities in an Eternally Inflating Universe}, 
{\sl Phys. Rev. Lett.} {\bf 81} (1998) 5501 

\bibitem{carr}
B. J. Carr and M. J. Rees, 
{\it The Anthropic Principle and the Structure of the Physical World}, 
{\sl Nature} {\bf 278} (1979) 611 

\bibitem{hogan}
C. J. Hogan, {\it Why the Universe is Just So}, 
{\sl Rev. Mod. Phys.} {\bf 72} (2000) 1149  

\bibitem{aguirre} 
A. Aguirre and M. Tegmark, 
{\it Multiple Universes, Cosmic Coincidences, and other Dark Matters}, 
{\sl J. Cosmol. Astropart. Phys.} {\bf 01} (2005) 003 

\bibitem{tegold} 
M. Tegmark, {\it What does Inflation Really Predict?}
{\sl J. Cosmol. Astropart. Phys.} {\bf 04} (2005) 001 

\bibitem{barnes}
L. A. Barnes, 
{\it The Fine-Tuning of the Universe for Intelligent Life}, 
{\sl Pub. Astron. Soc. Australia} {\bf 29} (2012) 529  

\bibitem{bartip}
J. D. Barrow and F. J. Tipler, 
{\sl The Anthropic Cosmological Principle}, Oxford Univ. Press (1986) 

\bibitem{weisskopf}
V. F. Weisskopf, {\it Of Atoms, Mountains, and Stars:}
{\it A study in qualitative physics}, {\sl Science} {\bf 187} (1975) 605 

\bibitem{quintana} 
E. V. Quintana, T. Barclay, S. N. Raymond, et al.,  
{\it An Earth-Sized Planet in the Habitable Zone of a Cool Star}, 
{\sl Science} {\bf 344} (2014) 277  

\bibitem{etaearth}
E. A. Petigura, A. W. Howard, and G. W. Marcy,
{\it Prevalence of Earth-size Planets orbiting Sun-like Stars},  
{\sl Proc. Nat. Acad. Sci.} {\bf 110} (2013) 19273  

\bibitem{gungott} 
J. E. Gunn and J. R. Gott III, 
{\it On the Infall of Matter Into Clusters of Galaxies}
{\it and Some Effects on Their Evolution}, 
{\sl Astrophys. J.} {\bf 176} (1972) 1

\bibitem{schecter} 
W. H. Press and P. Schecter, {\it Formation of Galaxies and Clusters}
{\it of Galaxies by Self-Similar Gravitational Condensation}, 
{\sl Astrophys. J.} {\bf 187} (1974) 425 

\bibitem{kolbturn} 
E. W. Kolb and M. S. Turner, {\sl The Early Universe}, 
Addison-Wesley (1990)

\bibitem{nagamine} 
K. Nagamine and A. Loeb, 
{\it Future Evolution of Nearby Large-scale Structures}
{\it in a Universe Dominated by a Cosmological Constant}, 
{\sl New Astron.} {\bf 8} (2003) 439 

\bibitem{busha2005}
M. T. Busha, A. E. Evrard, F. C. Adams, and R. H. Weschler,
{\it The Ultimate Halo Mass in a $\Lambda$CDM Universe}, 
{\sl Mon. Not. R. Astron. Soc.} {\bf 363} (2005) L11 

\bibitem{busha2007}
M. T. Busha, A. E. Evrard, and F. C. Adams, 
{\it The Asymptotic Form of Cosmic Structure:}
{\it Small-Scale Power and Accretion History},  
{\sl Astrophys. J.} {\bf 665} (2007) 1

\bibitem{nfw}
J. F. Navarro, C. M. Frenk, and S.D.M. White, 
{\it A Universal Density Profile from Hierarchical Clustering}, 
{\sl Astrophys. J.} {\bf 490} (1997) 493 

\bibitem{hernquist}
L. Hernquist, 
{\it An Analytical Model for Spherical Galaxies and Bulges}, 
{\sl Astrophys. J.} {\bf 356} (1990) 359 

\bibitem{kravtsov} 
M. Surhud, B. Diemer, and A. V. Kravtsov, arXiv:1504.05591 
(2015)  

\bibitem{cannizzo} 
S. Cannizzo and T. C. Hollister, 
{\it Cold Dissipationless Collapse of Spherical Systems:} 
{\it Sensitivity to the initial density law}, 
{\sl Astrophys. J.} {\bf 400} (1992) 58  

\bibitem{boily} 
C. M. Boily and E. Athanassoula, {\it On the Equilibrium Morphology}
{\it of Systems Drawn from Spherical Collapse Experiments}, 
{\sl Mon. Not. R. Astron. Soc.} {\bf 369} (2006) 608

\bibitem{bintrem} 
J. Binney and S. Tremaine, {\sl Galactic Dynamics},  
Princeton Univ. Press (2008) 

\bibitem{dyson} 
F. J. Dyson, 
{\it Time Without End: Physics and biology in an open universe}, 
{\sl Rev. Mod. Phys.} {\bf 51} (1979) 447  

\bibitem{al1997}
F. C. Adams and G. Laughlin, {\it A Dying Universe:}
{\it The long-term fate and evolution of astrophysical objects}, 
{\sl Rev. Mod. Phys.} {\bf 69} (1997) 337 

\bibitem{binney81}
J. Binney, {\it Resonant Excitation of Motion Perpendicular to Galactic Planes}, 
{\sl Mon. Not. R. Astron. Soc.} {\bf 196} (1981) 455 

\bibitem{aplus} 
F. C. Adams, A. M. Bloch, S. C. Butler, J. M. Druce, and J. A. Ketchum, 
{\it Orbital Instabilities in a Triaxial Cusp Potential}, 
{\sl Astrophys. J.} {\bf 670} (2007) 1027 

\bibitem{binmer} 
J. Binney and M. Merrifield, {\sl Galactic Astronomy}, 
Princeton Univ. Press (1998) 

\bibitem{frozen} 
G. Laughlin and F. C. Adams, {\it The Frozen Earth:}
{\it Binary Scattering Events and the Fate of the Solar System}, 
{\sl Icarus} {\bf 145} (2000) 614  

\bibitem{lunine} 
J. I. Lunine, {\sl Astrobiology: A Multidisciplinary Approach}, 
Pearson (2005)

\bibitem{scharf} 
C. A. Scharf, {\sl Extrasolar Planets and Astrobiology}, 
Univ. Science Books (2009)

\bibitem{al2001}
F. C. Adams and G. Laughlin,
{\it Constraints on the Birth Aggregate of the Solar System},  
{\sl Icarus} {\bf 150} (2001) 151

\bibitem{gdawg} 
G. Li and F. C. Adams, {\it Cross-sections for Planetary Systems}
{\it Interacting with Passing Stars and Binaries}, 
{\sl Mon. Not. R. Astron. Soc.} {\bf 448} (2015) 344 

\bibitem{ab2005}
F. C. Adams and A. M. Bloch, 
{\it Orbits in Extended Mass Distributions:} 
{\it General Results and the Spirographic Approximation}, 
{\sl Astrophys. J.} {\bf 629} (2005) 204  

\bibitem{penar} 
J. Pe{\~ n}arrubia, Y.-Z. Ma, M. G. Walker, and A. McConnachie,
{\it A Dynamical Model of the Local Cosmic Expansion}, 
{\sl Mon. Not. R. Astron. Soc.} {\bf 443} (2014) 2204

\bibitem{prada}   
F. Prada, A. A. Klypin, A. J. Cuesta, J. E. Betancort-Rijo, and J. Primack, 
{\it Halo Concentrations in the Standard $\Lambda$ Cold Dark Matter Cosmology},
{\sl Mon. Not. R. Astron. Soc.} {\bf 423} (2012) 3018 

\bibitem{leitherer} 
C. Leitherer et al.,
{\it Starburst99: Synthesis Models for Galaxies with Active Star Formation},  
{\sl Astrophys. J. Suppl.} {\bf 123} (1999) 3

\bibitem{schaye} 
J. Schaye, {\it Star Formation Thresholds and Galaxy Edges: Why and Where}, 
{\sl Astrophys. J.} {\bf 609} (2004) 667  

\bibitem{kasting} 
J. F. Kasting, D. P. Whitmire, and R. T. Reynolds, 
{\it Habitable Zones around Main Sequence Stars}, 
{\sl Icarus} {\bf 101} (1993) 108 

\bibitem{bhrees} 
M. J. Rees, {\it Black Hole Models for Active Galactic Nuclei}, 
{\sl Ann. Rev. Astron. Astrophys.} {\bf 22} (1983) 471  

\bibitem{gzone} 
G. Gonzalez, D. Brownlee, and P. Ward, 
{\it The Galactic Habitable Zone: Galactic Chemical Evolution}, 
{\sl Icarus} {\bf 152} (2001) 185  

\bibitem{thompson} 
T. A. Thompson, {\it Gas Giants in Hot Water:} 
{\it Inhibiting giant planet formation and planet}
{\it habitability in dense star clusters through cosmic time}, 
{\sl Mon. Not. R. Astron. Soc.} {\bf 431} (2013) 63 

\bibitem{hayashi} 
C. Hayashi, {\it Structure of the Solar Nebula,}
{\it Growth and Decay of Magnetic Fields and}
{\it Effects of Magnetic and Turbulent Viscosities on the Nebula}, 
{\sl Prog. Theor. Phys. Suppl.} {\bf 70} (1981) 35 

\bibitem{sasaki}
S. Sasaki, in {\sl Origin of the Earth}, eds. H. E. Newsom and
J. H. Jones, Oxford Univ. Press) (1990) p. 195  

\bibitem{ikoma} 
M. Ikoma and H. Genda, 
{\it Constraints on the Mass of a Habitable}
{\it Planet with Water of Nebular Origin}, 
{\sl Astrophys. J.} {\bf 648} (2006) 696  

\bibitem{harrison} 
E. R. Harrison, {\it Olbers' Paradox and the Background Radiation}
{\it Density in an Isotropic Homogeneous Universe}, 
{\sl Mon. Not. R. Astron. Soc.} {\bf 131} (1965) 1

\bibitem{wesson} 
P. S. Wesson, {\it Olbers' Paradox and the Spectral Intensity of the}
{\it Extragalactic Background Light}, {\sl Astrophys. J.} {\bf 367} (1991) 399 

\bibitem{chandra} S. Chandrasekar, 
{\sl An Introduction to the Study of Stellar Structure}, 
Univ. Chicago Press (1939) 

\bibitem{phillips}
A. C. Phillips, {\sl The Physics of Stars}, Wiley (1994)  

\bibitem{adams} 
F. C. Adams, {\it Stars in Other Universes:}
{\it Stellar structure with different fundamental constants},
{\sl J. Cosmol. Astropart. Phys.} {\bf 08} (2008) 010 

\bibitem{burost} 
A. S. Burrows and J. P. Ostriker, 
{\it Astronomical Reach of Fundamental Physics}, 
{\sl Proc. Nat. Acad. Sci.} {\bf 111} (2014) 2409  

\bibitem{barrow}
J. D. Barrow, {\it Varying Constants}, 
{\sl Roy. Soc. London Trans.} A {\bf 363} (2005) 2139 

\bibitem{uzan}
J.-P. Uzan, {\it The Fundamental Constants and their Variation:}
{ \it Observational and theoretical status}, 
{\sl Rev. Mod. Phys.} {\bf 75} (2003) 403 

\bibitem{heller}
R. Heller and J. Armstrong, {\it Superhabitable Worlds}, 
{\sl Astrobiology} {\bf 14} (2014) 50 

\end{thebibliography}
\end{document}